\documentclass[twocolumn,showpacs,superscriptaddress,prl]{revtex4-1}
\usepackage{graphicx}
\usepackage{xcolor}
\usepackage{amsmath,amsthm,amssymb}
\usepackage{braket}
\usepackage[colorlinks,citecolor=blue,linkcolor=red]{hyperref}
\usepackage{mhchem}
\usepackage{multirow}
\usepackage[normalem]{ulem}

\begin{document}
 
\title{Magnetism Enhanced Surface Bonding of O$_{2}$ on CoPt}

\author{Kevin Allen}
\affiliation{Theoretical Division, Los Alamos National Laboratory, Los Alamos, New Mexico 87545, USA}
\affiliation{Department of Physics and Astronomy, Rice University, Houston, Texas 77005, USA}
\affiliation{Rice Center for Quantum Materials, Rice University, Houston, Texas 77005, USA}

\author{Christopher Lane}
\email{laneca@lanl.gov}
\affiliation{Theoretical Division, Los Alamos National Laboratory, Los Alamos, New Mexico 87545, USA}

\author{Emilia Morosan}
\affiliation{Department of Physics and Astronomy, Rice University, Houston, Texas 77005, USA}
\affiliation{Rice Center for Quantum Materials, Rice University, Houston, Texas 77005, USA}

\author{Jian-Xin Zhu}
\email{jxzhu@lanl.gov}
\affiliation{Theoretical Division, Los Alamos National Laboratory, Los Alamos, New Mexico 87545, USA}

\date{\today} 
\begin{abstract}
For large-scale deployment and use of polymer electrolyte fuel cells, high-performance electrocatalysts with low platinum consumption are desirable. One promising strategy to meet this demand is to explore alternative materials that retain catalytic efficiency while introducing new mechanisms for performance enhacement. In this study, we investigate a ferromagnetic CoPt as a candidate material to accelerate oxygen reduction reactions. By using density functional theory calculations, we find the spin-polarized Co-$d$ states to enhance O$_2$ surface bonding due to local exchange splitting of Co-$d$ carriers at the Fermi level. Furthermore, O and O$_2$ adsorption and dissociation energies are found to be tuned by varying the thickness of the Pt layers. Our study gives insight into the role magnetism plays in the oxygen reduction reaction process and how magnetic ions may aid in the design of new advanced catalysts. 
\end{abstract}


\maketitle 

\section{Introduction}
The oxygen reduction reaction (ORR) process is critical in the development of energy conversion technologies such as fuel cells~\cite{zhang2008pem} and metal-air batteries~\cite{cheng2012metal}. Controlling the reduction process on the cathode presents significant challenges due to the slow kinetics involved in activating O$_2$, breaking the O-O bond, and removing oxides. These obstacles place high demands on the catalyst~\cite{wang2019review,nie2015recent}. Currently, platinum (Pt) based materials are the most effective catalysts for accelerating sluggish ORR kinetics. Given the high economic cost of Pt, finding alternative catalysts that reduce Pt usage or fully substitute it without compromising performance is essential~\cite{debe2012electrocatalyst,shao2016recent, setzler2016activity, yu2012review}.

One emerging approach to tune catalytic reactions involving paramagnetic species, such as O$_2$~\cite{ren2021spin,okada2003effect,abel2021ferromagnetic,kicinski2019enhancement,vensaus2024enhancement,zhang2020recent,wang2016effect,feng2023recent}, is through the introduction of magnetic elements. The application of an external magnetic field to paramagnetic catalysts has been shown to enhance ORR activity, improving electron transfer efficiency through the alignment of unpaired spins\cite{okada2003effect}. Notable examples include Co$_3$O$_4$ nanofiber composites\cite{zeng2018magnetic} and Fe/N/S-Co-doped carbon gels\cite{kicinski2019enhancement}, with studies reporting substantial performance gains under moderate magnetic fields \cite{garces2019direct,yan2021direct}. It is expected that the adsorption (dissociation) energies and ORR activity~\cite{bhattacharjee2016improved} in general may be tuned by modifying the local electronic structure of a catalyst's active sites via the magnetic ions in close proximity. A common framework for predicting catalytic activity on transition metal surfaces is the \textit{d}-band center model, introduced by Hammer and N{\o}rskov~\cite{hammer2000theoretical} that relates the \textit{d}-band center position to adsorbent binding strength. However, it becomes insufficient when spin-polarized \textit{d}-states are considered. Spin exchange between the adsorbent and the surface add complexity that the \textit{d}-band model does not fully capture, especially in systems with magnetic ions. 

Spin polarization can either enhance or diminish the ORR process, depending on the magnetic and electronic properties of the catalysts ~\cite{bhattacharjee2016improved,wang2016effect}. To better understand this interplay, it is crucial to study materials where magnetism significantly influences catalytic activity. CoPt is one such material, known for its hard ferromagnetic properties and magnetocrystalline anisotropy \cite{ariake2005magnetic,klemmer1995magnetic,hu2014structural,li2020structural}. Beyond its magnetic characteristics, CoPt also exhibits strong catalytic activity in ORR~\cite{xia2021high,pan2022ordered,guo2013fept,li2019hard}. These combined properties make CoPt an ideal candidate for exploring how magnetism can be harnessed to enhance ORR performance.

\begin{figure*}[ht!]
    \centering
    \includegraphics[width=0.9\textwidth]{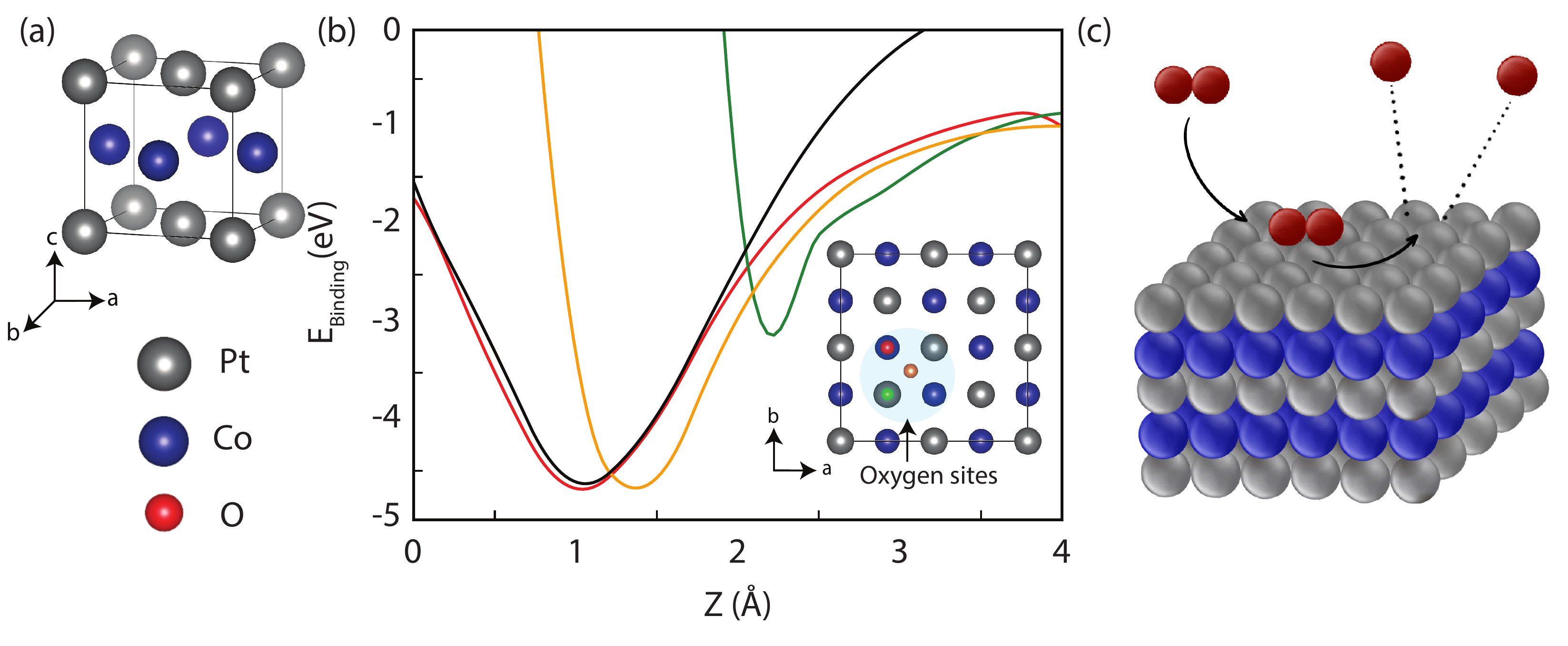}
    \caption{Crystal structure, binding energies, and pathway to oxygen reduction reaction for oxygen (O) on the (001) surface of L1$_{0}$-CoPt. (a) Crystal structure of bulk CoPt, with gray, blue and red spheres denoting the Platinum (Pt),Cobalt (Co), O atoms, respectively. (b) Binding energy as a function of O height from the slab surface for various surface binding sites, as indicated in the inset. The hollow (red) and bridge (orange) sites exhibit the strongest binding, which is 70 meV below the non-magnetic case (black).The top (green) site exhibits the weakest binding out of three. The height Z, is measured from the center of top layer. (c) Illustration for the O$_{2}$ adsorption and dissociation process on the (001) L1$_{0}$-CoPt surface.}
    \label{Fig1_background}
\end{figure*}

In this article, we perform a systematic first-principles study to investigate the adsorption and dissociation processes of O$_{2}$ on the CoPt (001) surface. Our results show that the intrinsic magnetic moment of Co enhances the binding strength. When considering the Pt layer thickness, we observe that a thin layer of Pt, which places Co atoms closer to the surface, significantly amplifies the surface binding effects. Finally, we explain the underlying mechanism for the enhanced binding energy by refining the \textit{d}-band center model to incorporate effects of spin exchange. Consequently, CoPt emerges as a promising material for surface-enhanced O$_2$ trapping, though it exhibits slower ORR kinetics compared to Pt due to the magnetism-induced increase in binding strength. Nonetheless, understanding how magnetism modifies adsorption opens the door to rationally tuning spin-related interactions. These include magnetic anisotropy, external fields, alloying or strain engineering to strike a balance between strong binding and efficient reaction kinetics.

\subsection{Methods}
Electronic structure calculations were carried out with density function theory (DFT) as implemented in the Vienna \textit{ab initio} simulation package (VASP) \cite{kresse1996efficient}. We used the pseudopotential projector augmented-wave method with an energy cutoff of 700 eV for the plane-wave basis set \cite{kresse1993ab}. Exchange-correlation effects were treated using Perdew-Burke-Ernzerhof (PBE) generalized gradient approximation density functional \cite{kresse1999ultrasoft, perdew1996generalized}. In order to simulate a surface, we used a $2 \times  2 \times 7$ slab for the (001) surface with a vacuum thickness of 20 \AA. We have chosen sufficiently large super cells to avoid interactions between the O atoms (O$_{2}$ molecules) in the neighboring periodic images. The Brillouin zone integration was performed using a $16 \times 16 \times 16$ ($8 \times 8 \times 1$) $\Gamma$-centered Monkhorst Pack \textit{k}-point grid for the bulk (surface) calculations. Spin-orbit coupling effects were included self-consistently. 

\section{Results}

To investigate how magnetism affects the ORR process, we concentrate on understanding the binding of O (O$_2$) to various surface sites on CoPt, as well as their adsorption and dissociation energies. Our study specifically examines L1$_{0}$-CoPt due its high structural stability, performance, and durability as an ORR catalyst~\cite{pan2022ordered, li2019hard,abel2021ferromagnetic}. Bulk CoPt crystallizes in the tetragonal L1$_{0}$ structure as shown in Fig.~\ref{Fig1_background} (a), where the gray, blue, and red spheres denote Pt, Co, and O atoms, respectively. We used the experimental unit cell with lattice parameters $a = 3.8$\,\AA\, and $c/a = 0.972$ in all calculations. DFT calculations were performed with the magnetic moments aligned along different crystalline axes: in-plane [100] and [110] directions, and along the out-of-plane [001] axis. In agreement with previous reports \cite{liu2016first}, we find a self-consistent magnetic moment of 1.959 $\mu_B$ for Co, and a 0.441 $\mu_B$ for Pt, with moments along the [001] (easy axis) direction yielding the lowest energy configuration.

\begin{figure*}[ht]
    \centering
    \includegraphics[width=0.9\textwidth]{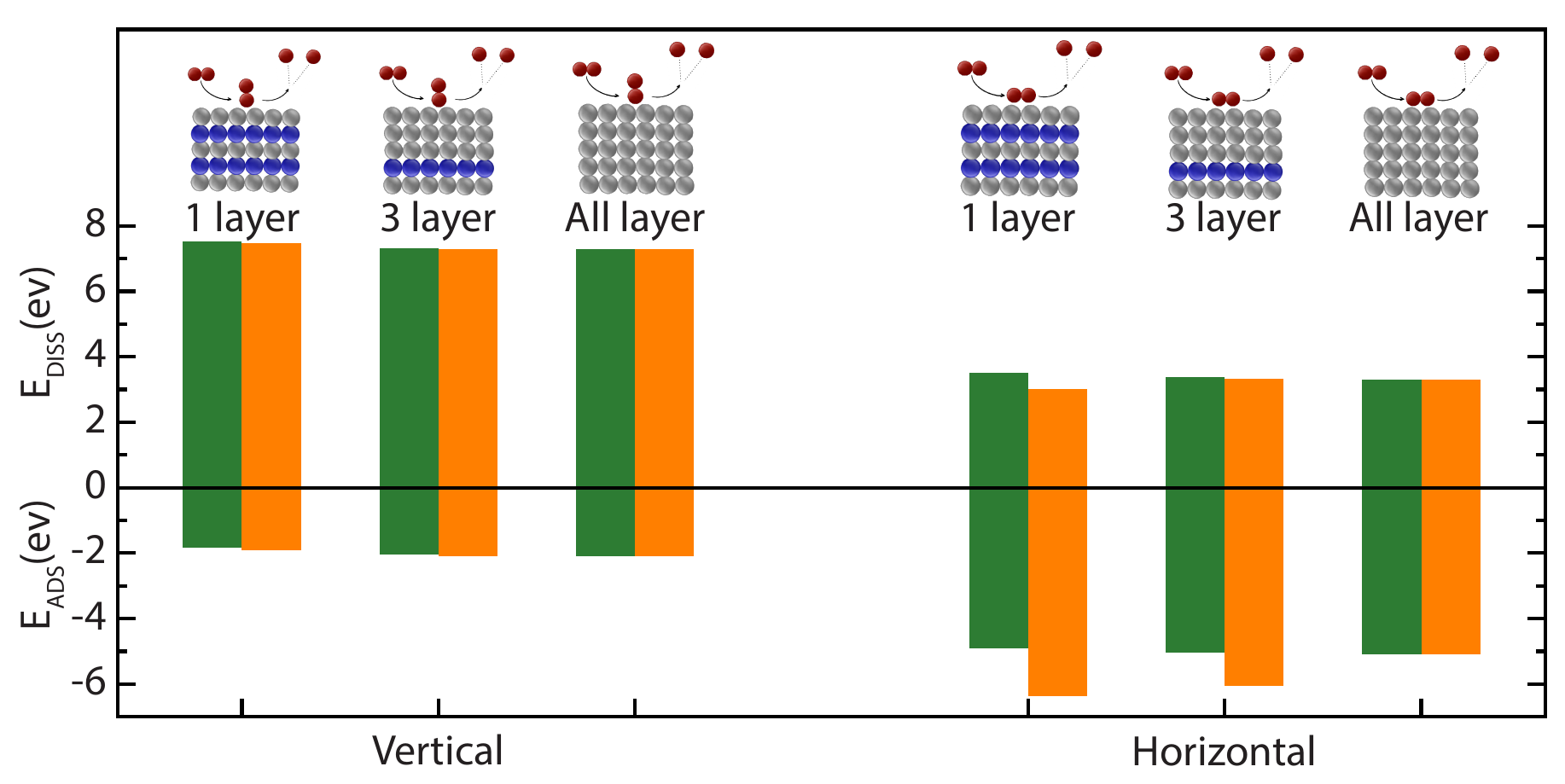}
    \caption{Comparison of the calculated adsorption and dissociation energies for one, three, and pure Pt surface layers, with (orange bars) and without (green bars) spin polarization on the cobalt atomc sites, and O$_{2}$ molecular orientations for a (001) surface of L1$_{0}$-CoPt.}
    \label{Fig2_Energies}
\end{figure*}

A key descriptor for catalytic activity is the binding energy between the catalyst and the adsorbent \cite{medford2015sabatier} at the various surface sites. Here, the [001] surface admits three unique sites including the top, bridge, and hollow sites, see the inset in Fig.~\ref{Fig1_background} (b). 
The binding energy of oxygen adsorbed on CoPt is defined as:
\begin{equation*}
\Delta E = E_{\mathrm{CoPt_{n}+O}} - E_{\mathrm{CoPt_{n}}} - E_{\mathrm{O}},
\end{equation*}
where $n$ denotes the number of surface Pt layers, E$_{\mathrm{CoPt_{n}+O}}$ is the total energy of the combined CoPt$+$O system, E$_{\mathrm{CoPt_{n}}}$ is the total energy of bare CoPt surface, and E$_{\mathrm{O}}$ is the energy of an isolated oxygen atom. 

Figure~\ref{Fig1_background} (b) presents the binding energy as a function of O height from the slab surface for various surface binding sites on the $L1_0-$CoPt (001) surface. The highly coordinated hollow site (red line) is the most energetically favorable, exhibiting a binding energy of $-4.69$ eV at a distance of $\approx 1.2$ \AA\ from the surface. The bridge site (orange line) is slightly higher in energy by 10 meV, suggesting possible mixing between these sites at room temperature. The top site, on the other hand, displays a significantly reduced binding energy of $-3.12$ eV, for which we anticipate accidental perturbations will drive O to migrate to the bridge and hollow sites. Notably, by setting the spin polarization to zero on the Co atomic sites in the calculations, we find the binding energy to reduce by 70 meV (black line). That is, the presence of magnetism appears to strengthen the CoPt-O bond. 

To further elucidate the role that Co plays in the ORR process, we examine the adsorption and dissociation of O$_{2}$ on the CoPt surface with increasing surface layers of Pt. Initially, O$_{2}$ binds to the surface with an energy of
\begin{equation*}
     \Delta E_{\mathrm{ADS}} = E_{\mathrm{CoPt_{n}+O_{2}}} - E_{\mathrm{CoPt_{n}}} - E_{\mathrm{O}_{2}},
\end{equation*}
similar to atomic oxygen, as illustrated in Fig.~\ref{Fig1_background}(c). Then, following the adsorption of O$_{2}$ the energy necessary to dissociate the O$_{2}$ molecule is
\begin{equation*}
     \Delta E_{\mathrm{DISS}} = E_{\mathrm{CoPt_{n}+O_{2}}} - E_{\mathrm{CoPt_{n}}} - 2E_{\mathrm{O}}.
\end{equation*}
The balance between these two energies dictates the kinetics of the ORR process.

\begin{figure*}
    \centering
    \includegraphics[width=0.9\textwidth]{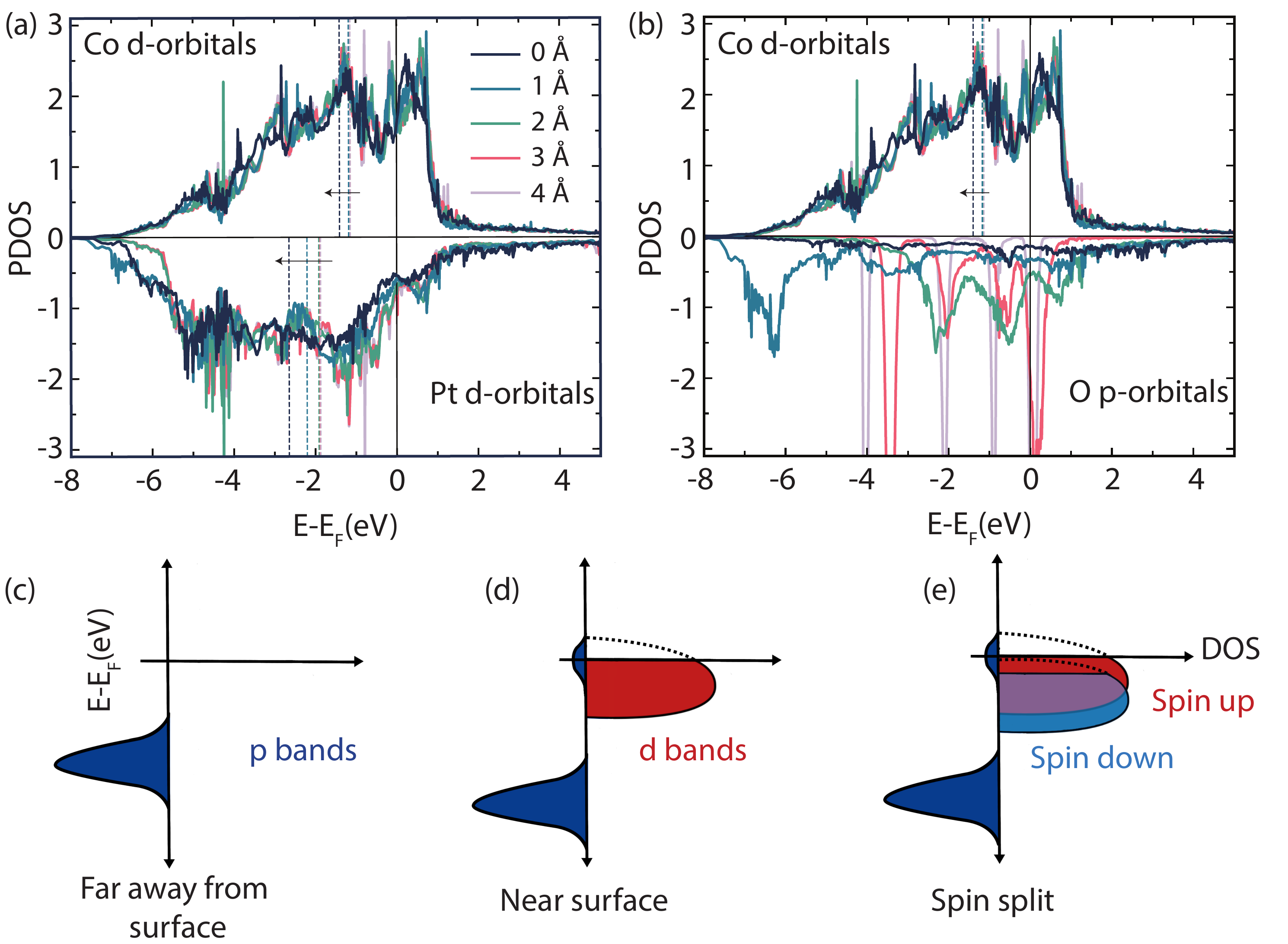}
    \caption{Comparison of spin polarized Co and Pt \textit{d}-band centers and the effect of magnetism on surface bonding. Atomic site resolved projected density of states for (a) Co-$d$ and Pt-$d$, and (b) Co-$d$ and O-$p$ states as a function of oxygen height from the L1$_{0}$-CoPt (001) surface. Schematic of the surface-adsorbent bonding for (c) a free oxygen atom, (d) oxygen bonding with a non-magnetic surface, and (e) the effect of spin exchange splitting on surface-adsorbent bonding.}
    \label{Fig3_Model}
\end{figure*}

Figure~\ref{Fig2_Energies} compares adsorption and dissociation energies for various surface Pt thicknesses, Co polarizations, and molecular orientations of O$_2$ for a (001) surface of CoPt. When Co is not spin polarized, increasing the number of Pt surface layers, monotonically decreases (increases) E$_{\mathrm{ADS}}$ (E$_{\mathrm{DISS}}$) to the pure Pt slab in accord with previously reported values \cite{gross2003unified,xue2018dissociative,li2016first,yang2010density,norskov2004origin,xu2004adsorption,kitchin2004modification}. However, when Co is allowed to be polarized, the trend inverts. That is, local exchange interactions between Co and O$_{2}$ induce a systematic decrease in E$_{\mathrm{ADS}}$ (E$_{\mathrm{DISS}}$); while when the Co layers go deeper away from the surface, the energies E$_{\mathrm{ADS}}$ and E$_{\mathrm{DISS}}$ approach to the pure Pt case. Overall the presence of magnetic Co atoms near the surface enhances the binding of O$_2$ to the surface of CoPt by $\approx$ 0.1 eV per O. Moreover, the orientation of the O$_{2}$ molecule has a marked effect on the adsorption and dissociation energies, yielding significantly reduced values when O$_{2}$ is horizontally oriented in almost all cases. This effect has also been reported for L1$_{0}$-FePt \cite{lu2020regulation}. We also found our results to be insensitive to the orientation of the Co magnetic moments, e.g. in or out of the plane.

The enhanced surface bonding in the presence of magnetism can be understood by examining the chemical bond effect between the CoPt surface and O$_{2}$. Bond formation between a transition metal and an adsorbent is often described by the \textit{d}-band model \cite{hammer2000theoretical}. The position and width of the \textit{d}-band relative to the Fermi level influences the strength of the chemical bond, and therefore, the catalytic activity of the surface. However, while the \textit{d}-band model provides valuable insights for non-magnetic systems, it is inadequate to fully capture the catalytic activity of magnetically polarized surfaces \cite{hammer2000theoretical,bhattacharjee2016improved}. Additionally, understanding how bond formation emerges as a function of adsorbent distance from the surface can further clarify the nature of the chemical bond \cite{norskov2011density,hammer2000theoretical}.

Figure ~\ref{Fig3_Model} (a) and (b) compares the atomic site resolved projected density of states for Co-$d$ and Pt-$d$, and Co-$d$ and O-$p$ orbitals, respectively, as a function of O height from the L1$_{0}$-CoPt (001) surface, where the vertical dashed lines indicate the \textit{d}-band center:
\begin{equation*}
     \epsilon_{d} = \frac{\int \epsilon g_d(\epsilon) d\epsilon}{\int g_d(\epsilon) d\epsilon};,
\end{equation*} 
where $g_d(\epsilon)$ is the Co-$d$ or Pt-$d$ partial density of states. The Co \textit{d}-band center is closer to the Fermi level by $\sim$1 eV for almost all O (O$_{2}$) distances from the surface, indicating Co is predominantly facilitating O (O$_{2}$) binding to the surface. Consequently, when Co is not spin polarized the \textit{d}-band center moves to the left, as indicated by the black arrow in Fig.~\ref{Fig3_Model} (a) and (b), suggesting weaker Co-O interactions, and thus weaker O surface binding CoPt. On comparing the Co-\textit{d} and O-\textit{p} density of states (Fig. \ref{Fig3_Model} (b)), they are found to overlap and evolve together as the O atom is brought closer to the surface, thereby suggesting strong hybridization between these states. In particular, when the O atom is far from the surface, the O-\textit{p} states are highly localized, but when oxygen is brought in proximity to the surface, the density of states becomes broader and splits, forming bonding and antibonding states.

To rationalize the effect of magnetism on chemical bonding at the surface, we extend the common \textit{d}-band 
picture \cite{bhattacharjee2016improved,hammer2000theoretical} as follows. An O adsorbent far away from the metal surface, i.e., 3 - 4 \AA, produces sharply peaked states below the Fermi level (Fig.~\ref{Fig3_Model} (c)). As the adsorbent approaches the surface its wave function starts to overlap and hybridize with that of the metallic surface. This broadens, shifts, and splits the adsorbent states into bonding and antibonding pairs (Fig.~\ref{Fig3_Model} (d)). The strength of the hybridization is gauged by the amount of \textit{d}-states available for bonding at the Fermi level, typically quantified by the proximity of the \textit{d}-band center to the Fermi energy \cite{norskov2004origin,nilsson2005electronic,greeley2002electronic}. If exchange splitting is present, as is the case for ferromagnetic CoPt, the \textit{d}-states are split, with the majority spin states pushed below the Fermi level rendering them inactive, whereas the minority spin states are shifted toward higher energies, thereby making more states available at the Fermi level for bonding and giving rise to stronger surface-adsorbent hybridization (Fig.~\ref{Fig3_Model} (e)). These minority-spin states near the Fermi level can accept electron density from the adsorbate, enhancing the orbital overlap and leading to stronger chemical bonding.

The activity of the oxygen reduction reaction (ORR) is highly sensitive to the \textit{d}-orbital configuration and charge transfer properties of the metallic surface \cite{kicinski2019enhancement}. To elucidate these interactions, we analyze the projected density of states (PDOS) and identify the key atomic orbitals involved in bonding. Specifically, the \textit{d}$_z^2$-orbitals of Co and Pt are found to be the primary contributors to the bonding between the adsorbent and the surface, while the $d_{xy}$- and $d_{yz}$-orbitals of Pt play comparatively minor roles. Furthermore, significant charge transfer is observed between the \textit{p}$_z$-states of oxygen and the Pt surface layers, consistent with the behavior expected of efficient catalysts such as Pt. However, when the Pt layer is reduced in thickness, cobalt assumes a dominant role in the bonding interaction. Due to the exchange splitting of its \textit{d}-states, the extent of charge transfer is diminished, which adversely affects the catalytic efficiency and leads to surface poisoning (i.e. partial or total deactivation of a catalyst caused by the very strong interaction of some reaction specifies with the active sites on the catalyst surface\cite{wandelt2018encyclopedia}).

In addition to the subtle balance between \textit{d}-band center and magnetic exchange splitting, surface-adsorbent interactions can be further enhanced when the adsorbent has an appreciable magnetic dipole \cite{bhattacharjee2016improved}. O is paramagnetic and, in light of its two unpaired electrons, the electrons can align with an external magnetic field, causing them to be attracted to the field \cite{okada2003effect}. Experimentally, the inclusion of an external magnetic field can, in principle, facilitate and increase oxygen transfer in the ORR process \cite{wang2016effect}. However, if the applied external field is lower than the saturation magnetization of a ferromagnetic catalyst, then performance can be degraded by reduced oxygen transfer rates. As a result, the inclusion of magnetism in the ORR process is a delicate process that depends sensitively on multiple competing factors that dictate catalyst performance. In systems like CoPt, where magnetism leads to overly strong binding, one strategy may be to modulate the magnetic exchange interaction through strain to weaken the binding just enough to optimize reaction kinetics. Alternatively, local field enhancement or selective surface engineering (adding several layers of Pt) could be used to spatially confine magnetic effects without universally increasing adsorption strength.

\section{Conclusion}
In summary, we have demonstrated that magnetism plays a critical role in influencing the ORR process of O$_2$ with L1$_{0}$-CoPt. Specifically, the presence of spin polarized cobalt atoms near the surface enhances bonding between O$_2$ and the CoPt, tipping the scales towards slower ORR kinetics. Our findings suggest that catalytic performance can be tuned by adjusting the magnetic and structural environment, such as increasing Pt overlayer thickness. This insight can guide the design of future catalysts leveraging magnetic properties to improve ORR efficiency and reduce reliance on platinum-based materials.

\section*{Acknowledgements}  
\begin{acknowledgments}
We acknowledge helpful discussions with Xiaojing Wang, Shengzhi Zhang, and Vivien Zapf. The work at Los Alamos National Laboratory was carried out under the auspices of the US Department of Energy (DOE) National Nuclear Security Administration under Contract No. 89233218CNA000001. It was supported by the LANL LDRD Program, and in part by the Center for Integrated Nanotechnologies, a DOE BES user facility, in partnership with the LANL Institutional Computing Program for computational resources. Additional computations were performed at the National Energy Research Scientific Computing Center (NERSC), a U.S. Department of Energy Office of Science User Facility located at Lawrence Berkeley National Laboratory, operated under Contract No. DE-AC02-05CH11231 using NERSC award ERCAP0020494. 
K.J.A. and E.M. have been supported by the Robert A. Welch Foundation under grant No. C-2114. 
\end{acknowledgments}

\bibliographystyle{naturemag}
\bibliography{CoPt_Refs}

\begin{thebibliography}{10}
\expandafter\ifx\csname url\endcsname\relax
  \def\url#1{\texttt{#1}}\fi
\expandafter\ifx\csname urlprefix\endcsname\relax\def\urlprefix{URL }\fi
\providecommand{\bibinfo}[2]{#2}
\providecommand{\eprint}[2][]{\url{#2}}

\bibitem{zhang2008pem}
\bibinfo{author}{Zhang, J.}
\newblock \emph{\bibinfo{title}{PEM fuel cell electrocatalysts and catalyst
  layers: fundamentals and applications}} (\bibinfo{publisher}{Springer Science
  \& Business Media}, \bibinfo{year}{2008}).

\bibitem{cheng2012metal}
\bibinfo{author}{Cheng, F.} \& \bibinfo{author}{Chen, J.}
\newblock \bibinfo{title}{Metal--air batteries: from oxygen reduction
  electrochemistry to cathode catalysts}.
\newblock \emph{\bibinfo{journal}{Chemical Society Reviews}}
  \textbf{\bibinfo{volume}{41}}, \bibinfo{pages}{2172--2192}
  (\bibinfo{year}{2012}).

\bibitem{wang2019review}
\bibinfo{author}{Wang, X.} \emph{et~al.}
\newblock \bibinfo{title}{Review of metal catalysts for oxygen reduction
  reaction: from nanoscale engineering to atomic design}.
\newblock \emph{\bibinfo{journal}{Chem}} \textbf{\bibinfo{volume}{5}},
  \bibinfo{pages}{1486--1511} (\bibinfo{year}{2019}).

\bibitem{nie2015recent}
\bibinfo{author}{Nie, Y.}, \bibinfo{author}{Li, L.} \& \bibinfo{author}{Wei,
  Z.}
\newblock \bibinfo{title}{Recent advancements in pt and pt-free catalysts for
  oxygen reduction reaction}.
\newblock \emph{\bibinfo{journal}{Chemical Society Reviews}}
  \textbf{\bibinfo{volume}{44}}, \bibinfo{pages}{2168--2201}
  (\bibinfo{year}{2015}).

\bibitem{debe2012electrocatalyst}
\bibinfo{author}{Debe, M.~K.}
\newblock \bibinfo{title}{Electrocatalyst approaches and challenges for
  automotive fuel cells}.
\newblock \emph{\bibinfo{journal}{Nature}} \textbf{\bibinfo{volume}{486}},
  \bibinfo{pages}{43--51} (\bibinfo{year}{2012}).

\bibitem{shao2016recent}
\bibinfo{author}{Shao, M.}, \bibinfo{author}{Chang, Q.},
  \bibinfo{author}{Dodelet, J.-P.} \& \bibinfo{author}{Chenitz, R.}
\newblock \bibinfo{title}{Recent advances in electrocatalysts for oxygen
  reduction reaction}.
\newblock \emph{\bibinfo{journal}{Chemical reviews}}
  \textbf{\bibinfo{volume}{116}}, \bibinfo{pages}{3594--3657}
  (\bibinfo{year}{2016}).

\bibitem{setzler2016activity}
\bibinfo{author}{Setzler, B.~P.}, \bibinfo{author}{Zhuang, Z.},
  \bibinfo{author}{Wittkopf, J.~A.} \& \bibinfo{author}{Yan, Y.}
\newblock \bibinfo{title}{Activity targets for nanostructured
  platinum-group-metal-free catalysts in hydroxide exchange membrane fuel
  cells}.
\newblock \emph{\bibinfo{journal}{Nature nanotechnology}}
  \textbf{\bibinfo{volume}{11}}, \bibinfo{pages}{1020--1025}
  (\bibinfo{year}{2016}).

\bibitem{yu2012review}
\bibinfo{author}{Yu, W.}, \bibinfo{author}{Porosoff, M.~D.} \&
  \bibinfo{author}{Chen, J.~G.}
\newblock \bibinfo{title}{Review of pt-based bimetallic catalysis: from model
  surfaces to supported catalysts}.
\newblock \emph{\bibinfo{journal}{Chemical reviews}}
  \textbf{\bibinfo{volume}{112}}, \bibinfo{pages}{5780--5817}
  (\bibinfo{year}{2012}).

\bibitem{ren2021spin}
\bibinfo{author}{Ren, X.} \emph{et~al.}
\newblock \bibinfo{title}{Spin-polarized oxygen evolution reaction under
  magnetic field}.
\newblock \emph{\bibinfo{journal}{Nature communications}}
  \textbf{\bibinfo{volume}{12}}, \bibinfo{pages}{2608} (\bibinfo{year}{2021}).

\bibitem{okada2003effect}
\bibinfo{author}{Okada, T.} \emph{et~al.}
\newblock \bibinfo{title}{The effect of magnetic field on the oxygen reduction
  reaction and its application in polymer electrolyte fuel cells}.
\newblock \emph{\bibinfo{journal}{Electrochimica acta}}
  \textbf{\bibinfo{volume}{48}}, \bibinfo{pages}{531--539}
  (\bibinfo{year}{2003}).

\bibitem{abel2021ferromagnetic}
\bibinfo{author}{Abel, F.~M.} \emph{et~al.}
\newblock \bibinfo{title}{Ferromagnetic l10-structured copt nanoparticles for
  permanent magnets and low pt-based catalysts}.
\newblock \emph{\bibinfo{journal}{ACS Applied Nano Materials}}
  \textbf{\bibinfo{volume}{4}}, \bibinfo{pages}{9231--9240}
  (\bibinfo{year}{2021}).

\bibitem{kicinski2019enhancement}
\bibinfo{author}{Kici{\'n}ski, W.} \emph{et~al.}
\newblock \bibinfo{title}{Enhancement of pgm-free oxygen reduction
  electrocatalyst performance for conventional and enzymatic fuel cells: The
  influence of an external magnetic field}.
\newblock \emph{\bibinfo{journal}{Applied Catalysis B: Environmental}}
  \textbf{\bibinfo{volume}{258}}, \bibinfo{pages}{117955}
  (\bibinfo{year}{2019}).

\bibitem{vensaus2024enhancement}
\bibinfo{author}{Vensaus, P.}, \bibinfo{author}{Liang, Y.},
  \bibinfo{author}{Ansermet, J.-P.}, \bibinfo{author}{Soler-Illia, G.~J.} \&
  \bibinfo{author}{Lingenfelder, M.}
\newblock \bibinfo{title}{Enhancement of electrocatalysis through magnetic
  field effects on mass transport}.
\newblock \emph{\bibinfo{journal}{nature communications}}
  \textbf{\bibinfo{volume}{15}}, \bibinfo{pages}{2867} (\bibinfo{year}{2024}).

\bibitem{zhang2020recent}
\bibinfo{author}{Zhang, Y.} \emph{et~al.}
\newblock \bibinfo{title}{Recent advances in magnetic field-enhanced
  electrocatalysis}.
\newblock \emph{\bibinfo{journal}{ACS Applied Energy Materials}}
  \textbf{\bibinfo{volume}{3}}, \bibinfo{pages}{10303--10316}
  (\bibinfo{year}{2020}).

\bibitem{wang2016effect}
\bibinfo{author}{Wang, L.} \emph{et~al.}
\newblock \bibinfo{title}{The effect of the internal magnetism of ferromagnetic
  catalysts on their catalytic activity toward oxygen reduction reaction under
  an external magnetic field}.
\newblock \emph{\bibinfo{journal}{Ionics}} \textbf{\bibinfo{volume}{22}},
  \bibinfo{pages}{2195--2202} (\bibinfo{year}{2016}).

\bibitem{feng2023recent}
\bibinfo{author}{Feng, Z.} \emph{et~al.}
\newblock \bibinfo{title}{Recent development of external magnetic field
  assisted oxygen evolution reaction-a mini review}.
\newblock \emph{\bibinfo{journal}{ChemCatChem}} \textbf{\bibinfo{volume}{15}},
  \bibinfo{pages}{e202300688} (\bibinfo{year}{2023}).

\bibitem{zeng2018magnetic}
\bibinfo{author}{Zeng, Z.} \emph{et~al.}
\newblock \bibinfo{title}{Magnetic field-enhanced 4-electron pathway for
  well-aligned co3o4/electrospun carbon nanofibers in the oxygen reduction
  reaction}.
\newblock \emph{\bibinfo{journal}{ChemSusChem}} \textbf{\bibinfo{volume}{11}},
  \bibinfo{pages}{580--588} (\bibinfo{year}{2018}).

\bibitem{garces2019direct}
\bibinfo{author}{Garc{\'e}s-Pineda, F.~A.}, \bibinfo{author}{Blasco-Ahicart,
  M.}, \bibinfo{author}{Nieto-Castro, D.}, \bibinfo{author}{L{\'o}pez, N.} \&
  \bibinfo{author}{Gal{\'a}n-Mascar{\'o}s, J.~R.}
\newblock \bibinfo{title}{Direct magnetic enhancement of electrocatalytic water
  oxidation in alkaline media}.
\newblock \emph{\bibinfo{journal}{Nature Energy}} \textbf{\bibinfo{volume}{4}},
  \bibinfo{pages}{519--525} (\bibinfo{year}{2019}).

\bibitem{yan2021direct}
\bibinfo{author}{Yan, J.} \emph{et~al.}
\newblock \bibinfo{title}{Direct magnetic reinforcement of electrocatalytic
  orr/oer with electromagnetic induction of magnetic catalysts}.
\newblock \emph{\bibinfo{journal}{Advanced Materials}}
  \textbf{\bibinfo{volume}{33}}, \bibinfo{pages}{2007525}
  (\bibinfo{year}{2021}).

\bibitem{bhattacharjee2016improved}
\bibinfo{author}{Bhattacharjee, S.}, \bibinfo{author}{Waghmare, U.~V.} \&
  \bibinfo{author}{Lee, S.-C.}
\newblock \bibinfo{title}{An improved d-band model of the catalytic activity of
  magnetic transition metal surfaces}.
\newblock \emph{\bibinfo{journal}{Scientific reports}}
  \textbf{\bibinfo{volume}{6}}, \bibinfo{pages}{35916} (\bibinfo{year}{2016}).

\bibitem{hammer2000theoretical}
\bibinfo{author}{Hammer, B.} \& \bibinfo{author}{N{\o}rskov, J.~K.}
\newblock \bibinfo{title}{Theoretical surface science and
  catalysis—calculations and concepts}.
\newblock In \emph{\bibinfo{booktitle}{Advances in catalysis}},
  vol.~\bibinfo{volume}{45}, \bibinfo{pages}{71--129}
  (\bibinfo{publisher}{Elsevier}, \bibinfo{year}{2000}).

\bibitem{ariake2005magnetic}
\bibinfo{author}{Ariake, J.}, \bibinfo{author}{Chiba, T.},
  \bibinfo{author}{Watanabe, S.}, \bibinfo{author}{Honda, N.} \&
  \bibinfo{author}{Ouchi, K.}
\newblock \bibinfo{title}{Magnetic and structural properties of co--pt
  perpendicular recording media with large magnetic anisotropy}.
\newblock \emph{\bibinfo{journal}{Journal of magnetism and magnetic materials}}
  \textbf{\bibinfo{volume}{287}}, \bibinfo{pages}{229--233}
  (\bibinfo{year}{2005}).

\bibitem{klemmer1995magnetic}
\bibinfo{author}{Klemmer, T.}, \bibinfo{author}{Hoydick, D.},
  \bibinfo{author}{Okumura, H.}, \bibinfo{author}{Zhang, B.} \&
  \bibinfo{author}{Soffa, W.}
\newblock \bibinfo{title}{Magnetic hardening and coercivity mechanisms in l10
  ordered fepd ferromagnets}.
\newblock \emph{\bibinfo{journal}{Scripta Metallurgica et Materialia}}
  \textbf{\bibinfo{volume}{33}}, \bibinfo{pages}{1793--1805}
  (\bibinfo{year}{1995}).

\bibitem{hu2014structural}
\bibinfo{author}{Hu, W.}, \bibinfo{author}{Yuan, H.}, \bibinfo{author}{Chen,
  H.}, \bibinfo{author}{Wang, G.} \& \bibinfo{author}{Zhang, G.}
\newblock \bibinfo{title}{Structural and magnetic properties of copt clusters}.
\newblock \emph{\bibinfo{journal}{Physics Letters A}}
  \textbf{\bibinfo{volume}{378}}, \bibinfo{pages}{198--206}
  (\bibinfo{year}{2014}).

\bibitem{li2020structural}
\bibinfo{author}{Li, L.}, \bibinfo{author}{Huang, R.}, \bibinfo{author}{Wen,
  Y.} \& \bibinfo{author}{Johnston, R.~L.}
\newblock \bibinfo{title}{Structural and magnetic properties of co-pt clusters:
  A spin-polarized density functional study}.
\newblock \emph{\bibinfo{journal}{Journal of Magnetism and Magnetic Materials}}
  \textbf{\bibinfo{volume}{503}}, \bibinfo{pages}{166651}
  (\bibinfo{year}{2020}).

\bibitem{xia2021high}
\bibinfo{author}{Xia, H.} \emph{et~al.}
\newblock \bibinfo{title}{High-efficient copt/activated functional carbon
  catalyst for li-o2 batteries}.
\newblock \emph{\bibinfo{journal}{Nano Energy}} \textbf{\bibinfo{volume}{84}},
  \bibinfo{pages}{105877} (\bibinfo{year}{2021}).

\bibitem{pan2022ordered}
\bibinfo{author}{Pan, Y.-T.} \emph{et~al.}
\newblock \bibinfo{title}{Ordered copt oxygen reduction catalyst with high
  performance and durability}.
\newblock \emph{\bibinfo{journal}{Chem Catalysis}}
  \textbf{\bibinfo{volume}{2}}, \bibinfo{pages}{3559--3572}
  (\bibinfo{year}{2022}).

\bibitem{guo2013fept}
\bibinfo{author}{Guo, S.} \emph{et~al.}
\newblock \bibinfo{title}{Fept and copt nanowires as efficient catalysts for
  the oxygen reduction reaction}.
\newblock \emph{\bibinfo{journal}{Angewandte Chemie}}
  \textbf{\bibinfo{volume}{125}}, \bibinfo{pages}{3549--3552}
  (\bibinfo{year}{2013}).

\bibitem{li2019hard}
\bibinfo{author}{Li, J.} \emph{et~al.}
\newblock \bibinfo{title}{Hard-magnet l10-copt nanoparticles advance fuel cell
  catalysis}.
\newblock \emph{\bibinfo{journal}{Joule}} \textbf{\bibinfo{volume}{3}},
  \bibinfo{pages}{124--135} (\bibinfo{year}{2019}).

\bibitem{kresse1996efficient}
\bibinfo{author}{Kresse, G.} \& \bibinfo{author}{Furthm{\"u}ller, J.}
\newblock \bibinfo{title}{Efficient iterative schemes for ab initio
  total-energy calculations using a plane-wave basis set}.
\newblock \emph{\bibinfo{journal}{Physical review B}}
  \textbf{\bibinfo{volume}{54}}, \bibinfo{pages}{11169} (\bibinfo{year}{1996}).

\bibitem{kresse1993ab}
\bibinfo{author}{Kresse, G.} \& \bibinfo{author}{Hafner, J.}
\newblock \bibinfo{title}{Ab initio molecular dynamics for open-shell
  transition metals}.
\newblock \emph{\bibinfo{journal}{Physical Review B}}
  \textbf{\bibinfo{volume}{48}}, \bibinfo{pages}{13115} (\bibinfo{year}{1993}).

\bibitem{kresse1999ultrasoft}
\bibinfo{author}{Kresse, G.} \& \bibinfo{author}{Joubert, D.}
\newblock \bibinfo{title}{From ultrasoft pseudopotentials to the projector
  augmented-wave method}.
\newblock \emph{\bibinfo{journal}{Physical review b}}
  \textbf{\bibinfo{volume}{59}}, \bibinfo{pages}{1758} (\bibinfo{year}{1999}).

\bibitem{perdew1996generalized}
\bibinfo{author}{Perdew, J.~P.}, \bibinfo{author}{Burke, K.} \&
  \bibinfo{author}{Ernzerhof, M.}
\newblock \bibinfo{title}{Generalized gradient approximation made simple}.
\newblock \emph{\bibinfo{journal}{Physical review letters}}
  \textbf{\bibinfo{volume}{77}}, \bibinfo{pages}{3865} (\bibinfo{year}{1996}).

\bibitem{liu2016first}
\bibinfo{author}{Liu, Z.}, \bibinfo{author}{Lei, Y.} \& \bibinfo{author}{Wang,
  G.}
\newblock \bibinfo{title}{First-principles computation of surface segregation
  in l10 copt magnetic nanoparticles}.
\newblock \emph{\bibinfo{journal}{Journal of Physics: Condensed Matter}}
  \textbf{\bibinfo{volume}{28}}, \bibinfo{pages}{266002}
  (\bibinfo{year}{2016}).

\bibitem{medford2015sabatier}
\bibinfo{author}{Medford, A.~J.} \emph{et~al.}
\newblock \bibinfo{title}{From the sabatier principle to a predictive theory of
  transition-metal heterogeneous catalysis}.
\newblock \emph{\bibinfo{journal}{Journal of Catalysis}}
  \textbf{\bibinfo{volume}{328}}, \bibinfo{pages}{36--42}
  (\bibinfo{year}{2015}).

\bibitem{gross2003unified}
\bibinfo{author}{Gro{\ss}, A.}, \bibinfo{author}{Eichler, A.},
  \bibinfo{author}{Hafner, J.}, \bibinfo{author}{Mehl, M.} \&
  \bibinfo{author}{Papaconstantopoulos, D.}
\newblock \bibinfo{title}{Unified picture of the molecular adsorption process:
  O2/pt (1 1 1)}.
\newblock \emph{\bibinfo{journal}{Surface science}}
  \textbf{\bibinfo{volume}{539}}, \bibinfo{pages}{L542--L548}
  (\bibinfo{year}{2003}).

\bibitem{xue2018dissociative}
\bibinfo{author}{Xue, T.}, \bibinfo{author}{Wu, C.}, \bibinfo{author}{Ding, X.}
  \& \bibinfo{author}{Sun, J.}
\newblock \bibinfo{title}{Dissociative adsorption of o 2 on strained pt (111)}.
\newblock \emph{\bibinfo{journal}{Physical Chemistry Chemical Physics}}
  \textbf{\bibinfo{volume}{20}}, \bibinfo{pages}{17927--17933}
  (\bibinfo{year}{2018}).

\bibitem{li2016first}
\bibinfo{author}{Li, R.}, \bibinfo{author}{Li, H.} \& \bibinfo{author}{Liu, J.}
\newblock \bibinfo{title}{First principles study of o2 dissociation on pt (111)
  surface: Stepwise mechanism}.
\newblock \emph{\bibinfo{journal}{International Journal of Quantum Chemistry}}
  \textbf{\bibinfo{volume}{116}}, \bibinfo{pages}{908--914}
  (\bibinfo{year}{2016}).

\bibitem{yang2010density}
\bibinfo{author}{Yang, Z.}, \bibinfo{author}{Wang, J.} \& \bibinfo{author}{Yu,
  X.}
\newblock \bibinfo{title}{Density functional theory studies on the adsorption,
  diffusion and dissociation of o2 on pt (111)}.
\newblock \emph{\bibinfo{journal}{Physics Letters A}}
  \textbf{\bibinfo{volume}{374}}, \bibinfo{pages}{4713--4717}
  (\bibinfo{year}{2010}).

\bibitem{norskov2004origin}
\bibinfo{author}{N{\o}rskov, J.~K.} \emph{et~al.}
\newblock \bibinfo{title}{Origin of the overpotential for oxygen reduction at a
  fuel-cell cathode}.
\newblock \emph{\bibinfo{journal}{The Journal of Physical Chemistry B}}
  \textbf{\bibinfo{volume}{108}}, \bibinfo{pages}{17886--17892}
  (\bibinfo{year}{2004}).

\bibitem{xu2004adsorption}
\bibinfo{author}{Xu, Y.}, \bibinfo{author}{Ruban, A.~V.} \&
  \bibinfo{author}{Mavrikakis, M.}
\newblock \bibinfo{title}{Adsorption and dissociation of o2 on pt- co and pt-
  fe alloys}.
\newblock \emph{\bibinfo{journal}{Journal of the American Chemical Society}}
  \textbf{\bibinfo{volume}{126}}, \bibinfo{pages}{4717--4725}
  (\bibinfo{year}{2004}).

\bibitem{kitchin2004modification}
\bibinfo{author}{Kitchin, J.}, \bibinfo{author}{N{\o}rskov, J.~K.},
  \bibinfo{author}{Barteau, M.} \& \bibinfo{author}{Chen, J.}
\newblock \bibinfo{title}{Modification of the surface electronic and chemical
  properties of pt (111) by subsurface 3d transition metals}.
\newblock \emph{\bibinfo{journal}{The Journal of chemical physics}}
  \textbf{\bibinfo{volume}{120}}, \bibinfo{pages}{10240--10246}
  (\bibinfo{year}{2004}).

\bibitem{lu2020regulation}
\bibinfo{author}{Lu, F.} \emph{et~al.}
\newblock \bibinfo{title}{Regulation of oxygen reduction reaction by the
  magnetic effect of l10-ptfe alloy}.
\newblock \emph{\bibinfo{journal}{Applied Catalysis B: Environmental}}
  \textbf{\bibinfo{volume}{278}}, \bibinfo{pages}{119332}
  (\bibinfo{year}{2020}).

\bibitem{norskov2011density}
\bibinfo{author}{N{\o}rskov, J.~K.}, \bibinfo{author}{Abild-Pedersen, F.},
  \bibinfo{author}{Studt, F.} \& \bibinfo{author}{Bligaard, T.}
\newblock \bibinfo{title}{Density functional theory in surface chemistry and
  catalysis}.
\newblock \emph{\bibinfo{journal}{Proceedings of the National Academy of
  Sciences}} \textbf{\bibinfo{volume}{108}}, \bibinfo{pages}{937--943}
  (\bibinfo{year}{2011}).

\bibitem{nilsson2005electronic}
\bibinfo{author}{Nilsson, A.} \emph{et~al.}
\newblock \bibinfo{title}{The electronic structure effect in heterogeneous
  catalysis}.
\newblock \emph{\bibinfo{journal}{Catalysis letters}}
  \textbf{\bibinfo{volume}{100}}, \bibinfo{pages}{111--114}
  (\bibinfo{year}{2005}).

\bibitem{greeley2002electronic}
\bibinfo{author}{Greeley, J.}, \bibinfo{author}{N{\o}rskov, J.~K.} \&
  \bibinfo{author}{Mavrikakis, M.}
\newblock \bibinfo{title}{Electronic structure and catalysis on metal
  surfaces}.
\newblock \emph{\bibinfo{journal}{Annual review of physical chemistry}}
  \textbf{\bibinfo{volume}{53}}, \bibinfo{pages}{319--348}
  (\bibinfo{year}{2002}).

\bibitem{wandelt2018encyclopedia}
\bibinfo{author}{Wandelt, K.}
\newblock \emph{\bibinfo{title}{Encyclopedia of interfacial chemistry: surface
  science and electrochemistry}} (\bibinfo{publisher}{Elsevier},
  \bibinfo{year}{2018}).

\end{thebibliography}

\appendix
\section{Adsorption and Dissociation Energies}

\begin{table}[h!]
\centering
\caption{Adsorption and dissociation energy values for the various configurations as shown in Fig.~\ref{Fig2_Energies}.}
\begin{tabular}{|c|c|c|c|c|}
\hline
\textbf{Energy} & \textbf{n Layers} & \textbf{Configuration} & \textbf{Vertical} & \textbf{Horizontal} \\ \hline
\multirow{6}{*}{\(E_\text{ADS}\)} 
& \multirow{2}{*}{1 Layer}  & With Moment     & -1.845487 & -4.911013 \\ \cline{3-5}
&                          & Without Moment  & -1.907118 & -6.357935 \\ \cline{2-5}
& \multirow{2}{*}{3 Layers} & With Moment     & -2.050356 & -5.045840 \\ \cline{3-5}
&                          & Without Moment  & -2.086599 & -6.044806 \\ \cline{2-5}
& \multirow{2}{*}{All Layers} & With Moment     & -2.088957 & -5.101708 \\ \cline{3-5}
&                          & Without Moment  & -2.088957 & -5.101708 \\ \hline
\multirow{6}{*}{\(E_\text{DISS}\)} 
& \multirow{2}{*}{1 Layer}  & With Moment     & 7.527343  & 3.504383  \\ \cline{3-5}
&                          & Without Moment  & 7.468168  & 3.017351  \\ \cline{2-5}
& \multirow{2}{*}{3 Layers} & With Moment     & 7.324932  & 3.369556  \\ \cline{3-5}
&                          & Without Moment  & 7.288688  & 3.330481  \\ \cline{2-5}
& \multirow{2}{*}{All Layers} & With Moment     & 7.286330  & 3.313688  \\ \cline{3-5}
&                          & Without Moment  & 7.286330  & 3.313688  \\ \hline
\end{tabular}
\label{tab:energy_values}
\end{table}

\clearpage

\end{document}